# Quantitative Pulse Shape-Instability Analysis Using 2D-Runs FROG

Pedram Abdolghader [1, 2], Rana Jafari [3], Abinash Das [3], Bilol Banerjee [4], Ellen P Crews [3], and Rick Trebino [3, *]

[1] Deutsches Elektronen-Synchrotron DESY, Notkestr. 85, 22607 Hamburg, Germany
[2] Institute of Quantum Optics, Leibniz University Hannover, Welfengarten 1, 30167 Hannover, Germany
[3] School of Physics, Georgia Institute of Technology, 837 State Street NW, Atlanta, GA 30332, USA
[4] Department of Statistics and Data Science, National University of Singapore, 6 Science Drive 2, Faculty of Science, NUS, Singapore 117546.

* Correspondence: rick.trebino@physics.gatech.edu

**Abstract:** We present a method for quantifying pulse-shape instability in a train of pulses using multi-shot Second-Harmonic-Generation Frequency-Resolved Optical Gating (SHG FROG). All versions of multi-shot FROG have previously shown the ability to distinguish stable from unstable pulse trains, as systematic differences appear between measured and retrieved traces when instability is present. This has proved possible because the recently introduced Retrieved-Amplitude N-grid Algorithmic (RANA) approach provides highly reliable pulse retrieval, even for unstable pulse trains and in the presence of noise, thus eliminating the possibility that algorithm stagnation, which mimics the effects of pulse-shape instability, could be confused for it. In other words, RANA's excellent performance ensures that any non-random discrepancies between measured and retrieved FROG traces reflect physical pulse-shape instability, rather than algorithmic stagnation. To begin to quantify such instability, we now introduce an instability parameter, $R$. It involves the use of the well-known statistical "Runs" test, which tests for systematic error in fits to one-dimensional (1D) data. A runs test counts the "runs"—consecutive points in the plot of the difference between the data and fit with the same sign (+ or -)—evaluating the goodness of the fit while minimizing the effects of random error. However, because FROG traces are functions of two variables, we must extend the usual 1D runs test to two dimensions, that is, to enumerate 2D runs—"hills" and "valleys" in the difference between measured and retrieved 2D FROG traces. Many small 2D runs indicate only random noise-like differences and hence a stable pulse train, whereas few large runs reflect additional systematic error and hence pulse-shape instability. But because random noise could contribute numerous meaningless runs in the wings of a FROG trace, we must also weight each hill and valley by its average measured trace value in order to minimize the effects of random noise. We show that $R$ is intuitive and reliable and, in addition, is trace-size–independent. As a result, it provides a clear metric of pulse-train stability vs. instability.

## 1. Introduction

Since the invention of the first lasers in the 1960s, techniques such as mode-locking [1–3], dispersion management [4–6], optical parametric chirped-pulse amplification (OPCPA) [7], and post-compression methods [8–12] have enabled the generation of ultrashort laser pulses with durations spanning from hundreds of nanoseconds down to just a few optical cycles. In parallel, advances in high-energy laser technology have made it possible to deliver pulses ranging from millijoules to hundreds of joules, with durations as short as a few hundred femtoseconds [13-22]. Recently developed high-peak-power laser systems now routinely achieve peak powers on the order of 10 PW per pulse, achieving irradiances exceeding $10^{23}$ W/cm² at focus [23]. Such energetic ultrashort pulses have enabled transformative applications across science and technology, most notably the emergence of laser-plasma accelerators, which employ intense femtosecond laser pulses to drive plasma waves capable of accelerating charged particles to near-light speeds over distances of only a few millimeters to centimeters. The idea was first proposed by Tajima and Dawson in 1979 [24] and became experimentally viable after the development of chirped-pulse amplification [25], which enabled the generation of sufficiently intense femtosecond pulses. Since the particle acceleration is highly sensitive to laser parameters, maintaining a stable and reproducible pulse train is essential to produce consistent, high-quality charged particle beams [26]. These secondary sources generate energetic photons [27], neutrons [28] and charged particles [29-34], with applications ranging from cancer proton therapy [32], phase contrast imaging using laser driven $K_\alpha$ X-ray sources [35], to neutron imaging [36], soft X-ray phase-contrast tomography [37], high-resolution MeV x-ray tomography [38], and fast ignition schemes in inertial confinement fusion (ICF) [39-40].

In high-power regimes, whether the application involves multi-shot averaging or not, pulse-shape stability is critically important, as even small fluctuations or a slight temporal delay of a pre-pulse relative to the main pulse can severely perturb or destroy the target under study. Pulse-shape stability is equally critical in fundamental research areas such as high-harmonic generation and attosecond light sources [41], extreme ultraviolet switching [42], attosecond optical switching [43], ultrafast optoelectronics [44], attosecond spectroscopy [45-46], and high-harmonic generation in gas and solid phase [47], all of which require precise control over both the temporal intensity and phase of the driving pulses. Similarly, in supercontinuum generation using hollow-core fibers or photonic crystal fibers, the stability of the supercontinuum strongly depends on the pump laser characteristics [48-49]. In nonlinear-optical microscopy applications, particularly those involving raster-scanned imaging, pulse-to-pulse variations can significantly degrade image quality. This is particularly critical in techniques like Optical Coherence Tomography and Stimulated Raman Scattering microscopy, where instability in the laser pulses can lead to image distortions or spectral artifacts [50-53].

Although major advances in laser science and engineering have made it possible to generate stable amplified ultrashort pulses with extreme peak and average powers at high repetition rate [54-55] and even pulse durations as short as a single optical cycle, ensuring pulse-shape stability remains a significant challenge. Even state-of-the-art systems can exhibit pulse-to-pulse fluctuations in intensity and phase vs. time and frequency that severely complicate experiments using them. Such instabilities can arise from a variety of sources, including pump-laser instability, thermal fluctuations, imperfect mode-locking, and even air turbulence in the laser beam path.

Since the early days of ultrafast optics, shot-to-shot variations in pulse shape, in both intensity and phase, have presented a major challenge for devices that measure ultrashort pulses [56-57]. When presented with a train of unstable pulses, intensity autocorrelation, the earliest method for measuring ultrashort pulses, displays a narrow "spike" atop a broad background, representing the coherent (repeatable) portion of the pulse train's intensity [57]. This spike is always narrower than the average pulse in the unstable train and generally indicates the shortest temporal spikes within more complex pulses. While it is generally a reliable indicator of pulse-intensity instability, this spike can be misleading: when instability is large, the broad background may be overlooked, and when instability is small, but nonzero, the spike blends into the more meaningful background, also reducing the width of the autocorrelation trace and making the pulse appear shorter than it actually is. Both cases can lead to the erroneous conclusion of a more stable and shorter pulse than is in fact present.

More recent, typically interferometric, techniques that measure the spectral phase can be even more problematic. This is because long complex pulses typically have complex spectral phases, while short simple pulses have a flat spectral phase. Indeed, for a given spectrum, the shortest pulse corresponds to a flat spectral phase. So, averaging the spectral phase over many different complex pulses yields an artificially flatter (or even perfectly flat) measured average spectral phase, misleadingly suggesting a shorter, often even transform-limited pulse. In other words, the average spectral phase is the frequency-domain description of the *coherent artifact* (a simple fact that is unfortunately not widely known). As a result, techniques that measure the average spectral phase, such as SPIDER, measure *only* the coherent artifact [59]. So, such interferometric techniques can only be trusted for a perfectly stable pulse train, but, as no device exists to confirm this latter fact, such measurements are prone to yielding erroneously short pulses.

Single-shot measurements could, in principle, resolve pulse-to-pulse variations. For example, a recent study at Lawrence Berkeley National Laboratory (LBNL) demonstrated such an approach for measuring pulses from a 100-TW Ti:Sapphire laser capable of delivering 2.5-J, sub-40-fs pulses at 1 Hz. In that work, single-shot complete intensity-and-phase measurements and correction were performed using the GRENOUILLE technique (a simple single-shot version of FROG) to help stabilize their laser-plasma accelerator [26]. But such single-shot measurements are only possible for amplified systems and are not usually possible in unamplified lasers due to insufficient pulse energy and also their high rep rates.

In fact, multi-shot averaging is typically employed in most practical pulse measurements. Fortunately, FROG and its variations have been known to provide an indicator of pulse-shape instability for over two decades [58]. Indeed, the first observation of this fact was quite dramatic, involving ultrabroadband supercontinuum pulses, initially thought to have smooth and stable spectra, but which were, in fact, shown using FROG to instead have extremely complex and chaotic spectra [58]. For unstable pulse trains, the measured FROG trace, an average over the many pulses in the measurement, no longer corresponds to that of any single pulse in the train, whereas FROG algorithms can (and should) only retrieve a trace corresponding to a single pulse. As a result, discrepancies between the measured and retrieved FROG traces reveal pulse-shape instability.

There has been one complication, however. Even the well-established standard generalized projections (GP) FROG pulse-retrieval algorithm may stagnate, especially for complex pulses, and, because stagnation also yields such discrepancies, these two unrelated effects could be confused for each other [60]. Fortunately, the recently developed, improved algorithmic approach, Retrieved-Amplitude N-grid Algorithmic (RANA), has proven extremely reliable and so eliminates the possibility of stagnation and hence this possible confusion. RANA's main innovation is to simply retrieve the pulse's approximate spectrum directly from the measured FROG trace and then use this recovered spectrum to construct vastly improved initial guesses for the electric field, which can then be used with any phase-retrieval algorithm, such as GP. RANA (using GP) has proven 100% reliable in retrieving stable, extremely complex pulses with time-bandwidth products (TBP) up to 100, even in presence of high noise and has been tested on sets of thousands of FROG traces [61-63]. Even for unstable pulse trains, it reliably converges to the field whose trace best matches the measured trace, ensuring that any difference reflects true pulse-shape instability rather than algorithm stagnation. Additionally, although the retrieved pulse field exhibits some smoothing of the pulse intensity vs. time, it exhibits a pulse duration and TBP that closely match the average values of the unstable pulses in the train, making it the best available representation of a typical pulse within the ensemble [64-65].

As a result, when retrieving the pulse from a FROG trace using RANA, algorithm stagnation is no longer a problem, and any discrepancies between the measured and retrieved traces are due to the usual random noise and possible systematic error due entirely to pulse-shape instability. This has also been demonstrated previously. [64-65]

The next challenge is to *quantify* the pulse-shape instability, rather than merely indicating its presence by eye. So, in this work, we have developed a parameter to do so. Our approach involves quantifying the systematic error in the difference between the measured and retrieved FROG traces. It is an extension of a well-known statistical approach for quantifying goodness of fit in one-dimensional data sets, called the "one-dimensional (1D) runs test" (or the Wald–Wolfowitz Runs Test, after its inventors) [66], which is commonly used to evaluate how well a model fits 1D experimental data, independent of random noise. Its use for a simple example data set is shown in Fig. 1a. A runs test is a reliable non-parametric statistical test used to check whether the difference between observed data points and a fit is random or there is some pattern in their ordering (indicating the presence of systematic error). In the difference data, a "run" is defined as a sequence of consecutive data points that share the same sign. For a good fit, there should be many short runs due to frequent sign changes in the difference data due only to random error. This behavior confirms that the model has successfully reproduced the main features of the data, and that the remaining differences are merely due to random noise. In contrast, when the differences maintain the same sign over long stretches of consecutive data points, that is, there are only a few runs, then the model consistently overestimates or underestimates the data, providing conclusive evidence of systematic error (in other words, a poor fit).

A FROG trace inherently involves a two-dimensional (2D) data trace, however, so any runs-based diagnostic for it must be extended to 2D. Multidimensional generalizations of runs tests have been developed and have been reported in the statistics literature, beginning with the seminal work of Friedman and Rafsky [67]. Their approach replaced the linear ordering of observations by a graph-theoretic construction. A more robust graph-based method was later introduced by Biswas *et al.* [68]. Unfortunately, neither approach is directly applicable to FROG traces due to the underlying geometry of the FROG's regular grid graphs. However, motivated by these ideas, we introduce a variation on their notion of 2D runs, tailored specifically to regular grid graphs, which naturally respects the geometry of FROG data.

We define our 2D runs statistic by first forming the pointwise difference between the measured and retrieved FROG traces (See Fig. 1b) for every grid point. This difference is itself a two-dimensional trace, and 2D runs are defined as connected sets of points of the grid over which the sign of the difference remains the same (+ or -). We then count the number of such connected 2D regions in the resulting sign graph. Visually, these regions can be interpreted as "hills" and "valleys" in the 2D difference-trace landscape.

When the retrieved trace faithfully reproduces the measured trace, the difference trace appears like random noise, with frequent sign-changes within consecutive grid points that fragment the landscape into many small regions (hills and valleys). Conversely, systematic error manifests as extended patches of the same sign, yielding hills and valleys of larger areas and fewer of them. The resulting statistic therefore provides a geometrically intuitive and distribution-free ("non-parametric") measure of structured mismatch between measured and retrieved FROG traces.

It is worth noting that, in general, a very large number of runs, whether 1D or 2D, while seemingly desirable, does not always necessarily mean that a model is meaningful. Such a case often implies "overfitting," that is, the use of too many parameters in the fitting function. In practice, a realistic good fit is one with an intermediate number of runs. However, overfitting is not relevant in FROG, where the model is known and fixed in advance and does not allow for variations in the number of parameters. So, in FROG, the more runs the better the fit and hence the more stable the pulse train. Fig. 1c shows the 2D runs in a sample difference trace corresponding to a lack of systematic error, while Fig. 1d shows the runs for a difference sample trace with considerable systematic error.

Another important issue in FROG is that the edge pixels of a FROG trace necessarily carry much less signal intensity and so contribute much less information to the trace. As a result, their contributed runs are numerous and dominated by even small amounts of random noise. To avoid these mostly meaningless contributions, we introduce a *weighting* approach to emphasize the more significant (central) higher-intensity regions of the FROG trace. Each 2D run is therefore weighted by the average intensity of the measured FROG trace for points in that run. This measure then considers only the statistically meaningful runs while minimizing the influence of low-signal areas or trace wings, where noise dominates the geography of the 2D runs. Figures 1e and f show sample traces of weighted 2D runs for cases of stable and unstable pulse trains for pulses with an average Gaussian shape. Note that the values in the difference-trace wings are near zero due to their low weights, as desired.

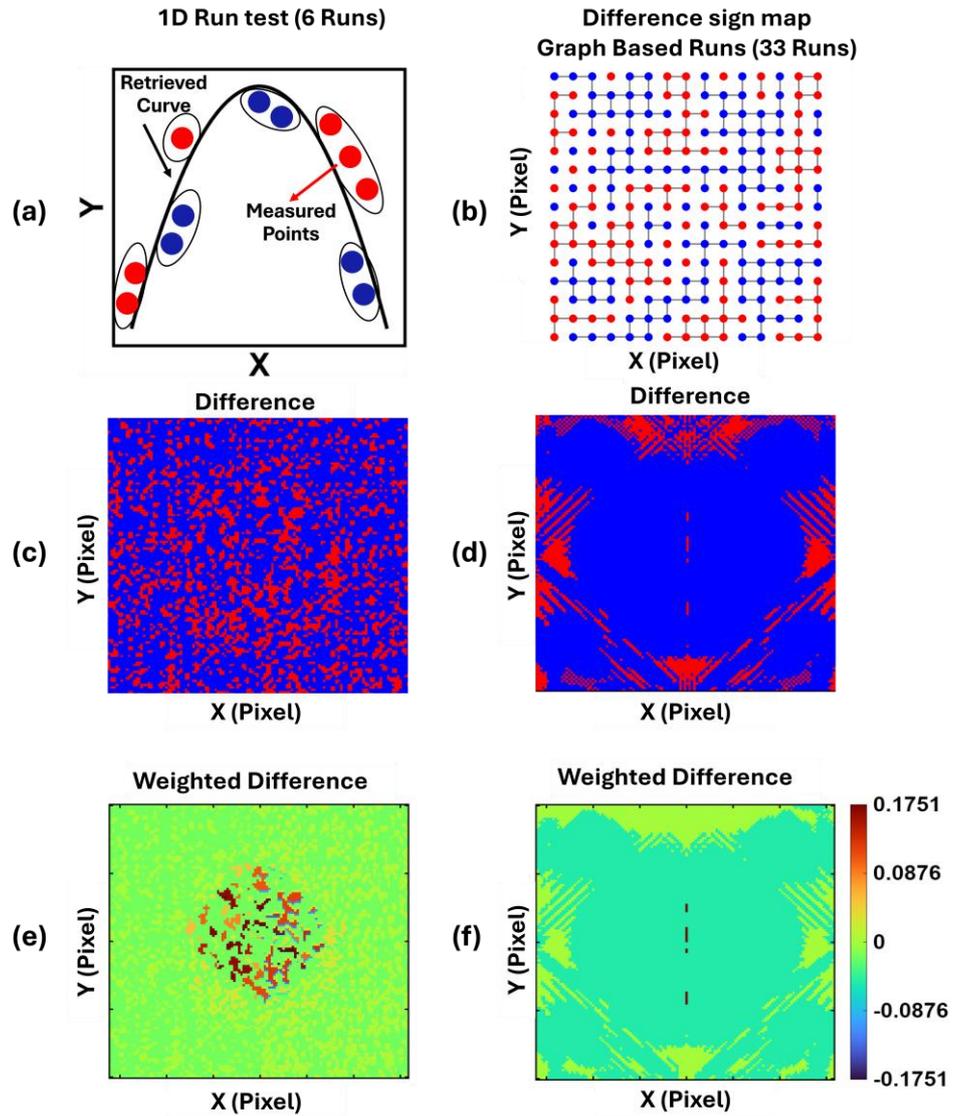

**Fig. 1.** (a) Example of 1D runs in curve fitting, where the difference curve contains 6 runs, with red dots indicating data points greater than the fit and blue dots indicating data point less than the fit. (We will use this convention in Figs. (b-d) also). (b) Graph-based two-dimensional runs test applied to a sample $15 \times 15$ difference trace with randomly distributed + and − pixels, as defined in (a). Line segments connect adjacent points of the same sign. A 2D run is a region of such adjacent points of the same sign isolated by points of opposite sign. (c) Example of such a graph-based two-dimensional runs test applied to the difference between measured and reconstructed FROG traces ($128 \times 128$ pixels) for a stable pulse train. Graph-based run-counting is performed by connecting neighboring pixels with identical signs, thereby forming distinct 2D runs. (d) Corresponding example with significant systematic error due an unstable pulse train. The difference trace exhibits a small number of large, connected regions (especially blue), resulting in fewer but significantly larger runs. (e) Corresponding example of *weighted* 2D runs for a trace with minimal systematic error and hence a stable pulse train, assuming a 2D Gaussian FROG trace. The difference trace exhibits a large number of small connected regions. (f) Corresponding example of *weighted* 2D runs for a trace with large systematic error and hence corresponding to an unstable pulse train, assuming 2D Gaussian weights, showing fewer but significantly larger runs. Note that runs in the wings, where the data trace is near zero contribute very little.

Applying this weighted 2D runs approach to the difference trace enables reliable identification of systematic errors in FROG difference traces indicative of the degree of pulse-shape instability. To quantify this effect in a trace-size-independent manner, we introduce (in the next section) a parameter that we call *R*, a scalar metric that directly measures the amount of systematic error in the difference trace and hence the pulse-shape instability.

To better illustrate the concept of our 2D weighted-runs approach, interpreted as a landscape of weighted "valleys" and "hills", we can use a three-dimensional visualization of the signed weighted map, as shown in Fig 2. In this representation, the difference trace of a stable pulse train with a TBP of 10 is plotted as a surface over delay and frequency, where the height corresponds to the weighted run value and the color indicates its magnitude and sign. To better illustrate the structure, a semi-transparent projection of the same map is added onto the bottom plane, allowing the underlying 2D distribution to be seen together with the 3D surface.

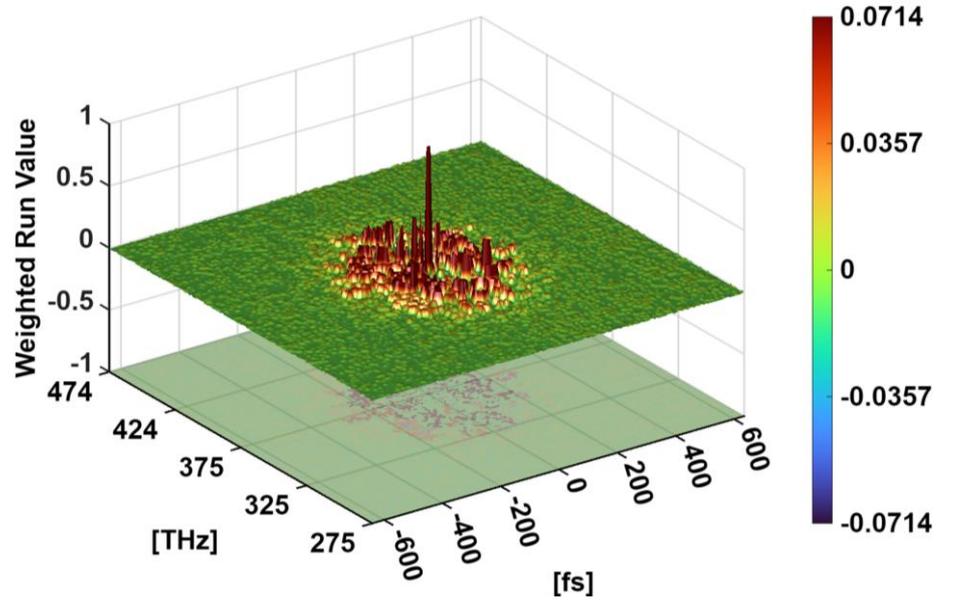

Fig. 2. A 3D visualization of the weighted map for the difference trace of a stable pulse train with a TBP of 10. The surface height represents the weighted run value, while the color encodes its magnitude and sign. A semi-transparent projection on the bottom plane allows the underlying 2D structure to be seen simultaneously, highlighting regions of higher and lower weighted contributions.

## 2. Graph-Based Method for 2D Run Analysis

We begin by defining the difference trace, $\Delta_{ij}$, which captures pixel-by-pixel discrepancies between the measured and retrieved traces, where the sign of each element indicates whether the retrieved intensity overestimates or underestimates the measured intensity locally. Measured FROG traces are assumed to be normalized to have a peak value of 1. We then compute 2D runs as described previously.

If the FROG trace is an $N \times N$ array, let the difference trace be:

$$\Delta_{ij} = I_{FROG}^{meas}(\omega_i, \tau_j) - I_{FROG}^{retr}(\omega_i, \tau_j) \quad \text{for } i,j \in \{1,\dots,N\},$$

Let *K* be the total number of 2D runs, *R(k)* be the set of pixels in the $k^{th}$ run, which we define to have $R_k$ pixels. As a result, the $k^{th}$ run will have a mean measured intensity:

$$\bar{I}_k = \frac{1}{|R_k|} \sum_{(i,j) \in R(k)} I_{FROG}^{meas}(\omega_i, \tau_j),$$

and the normalized weighted runs statistic, $R$, is

$$R = \frac{\sum_{k=1}^{K} \bar{I}_k}{\sum_{i,j} I_{FROG}^{meas}(\omega_i, \tau_j)}.$$

In the unrealistically ideal case in which the sign alternates from each pixel to the next (like a checkerboard), there will be $K = N^2$ runs, comprising one point each, and each run's weight will be the measured trace's value at the point. Both the numerator and denominator of $R$ will then have $N^2$ terms, each equaling a measured-trace data point, $I_{FROG}^{meas}(\omega_i, \tau_j)$. As a result, in this case, $R = 1$. In the worst-case scenario of an extremely poor fit (also unrealistic), where all measured points are larger (or smaller) than retrieved ones, there will be only one 2D run ($K = 1$), yielding $R_1 = N^2$ and $\bar{I}_k = \left(\frac{1}{N^2}\right) \Sigma\, I_{FROG}^{meas}$, which yields $R = 1/N^2$. Thus, $R$ ranges from $1/N^2$ to 1, and the more unstable the pulse shape, the lower the $R$ value.

Finally, $R$ is trace-size independent because both numerator and denominator scale in the same way when the trace size changes. In other words, when the measured and retrieved traces are resized (for example, by interpolation or by acquiring data with a higher sampling rate), the total number of pixels can increase or decrease, provided that the discrete-Fourier transform condition remains satisfied for the delay and frequency axes. This means that both the number of pixels contributing to each run and the sum of all pixel weights increase or decrease proportionally. This independence of $R$ from the trace size is shown in more detail in Appendix A.

## 3. Simulation Details

We investigated the performance of the 2D runs analysis using three different pairs of stable and unstable pulse trains. For consistency with our prior SHG FROG simulations (in which the pulse instability was qualitatively identified by visually observing the difference trace for systematic error), we used the same sets of stable and unstable test pulses: three stable pulse trains, each comprising a single pulse, and three unstable pulse trains, each comprising 5000 randomly varying pulses of the same average pulse length and time–bandwidth products (TBP) as the stable trains [59,64-65]. The pairs of trains had TBPs of 2.5, 5.0, and 10, respectively. Each pulse train was constructed by adding to a shorter stable short pulse a longer, unstable component, thereby introducing controlled pulse shape instability in the unstable trains. The stable pulse component in each pulse had a temporal full width at half maximum (FWHM) of 12 fs in all cases. Since the random components were necessarily longer, they were assigned higher energies before being added to the stable pulses, resulting in average temporal FWHMs of 26 fs, 54 fs, and 108 fs for both the stable and unstable trains. The pulse energies in the unstable trains were tailored to follow a normal distribution with a coefficient of variation of 10% (i.e., standard deviation over the mean).

Multi-shot SHG FROG traces for the stable pulse trains were computed for the single pulse comprising it since every pulse in it, by definition, yielded the same trace. Multi-shot SHG FROG traces for the unstable pulse trains were generated by averaging the individual SHG FROG traces of all the pulses within each train. Additionally, in all cases, 3% additive and 5% multiplicative noise were applied to all the traces. Finally, before the retrieval process, the traces were preprocessed using the same preprocessing techniques as are always performed when retrieving pulses from experimental FROG traces [60]. For the pulse trains with average temporal FWHMs of 26, 54, and 108 fs, we used trace sizes of 64 × 64, 128 × 128, and 256 × 256, respectively, ensuring that the intensities at the trace perimeters were less than $10^{-4}$ of peak of the trace to satisfy the FROG sampling rate [69]. We then retrieved pulses from the six resulting traces using the RANA approach [61-63], which incorporated the GP algorithm [60]. Fig. 3 illustrates sample pulses in the time domain from the three unstable pulse trains. For the stable pulse-train cases, each pulse in the train is identical. Each SHG FROG trace was analyzed individually.

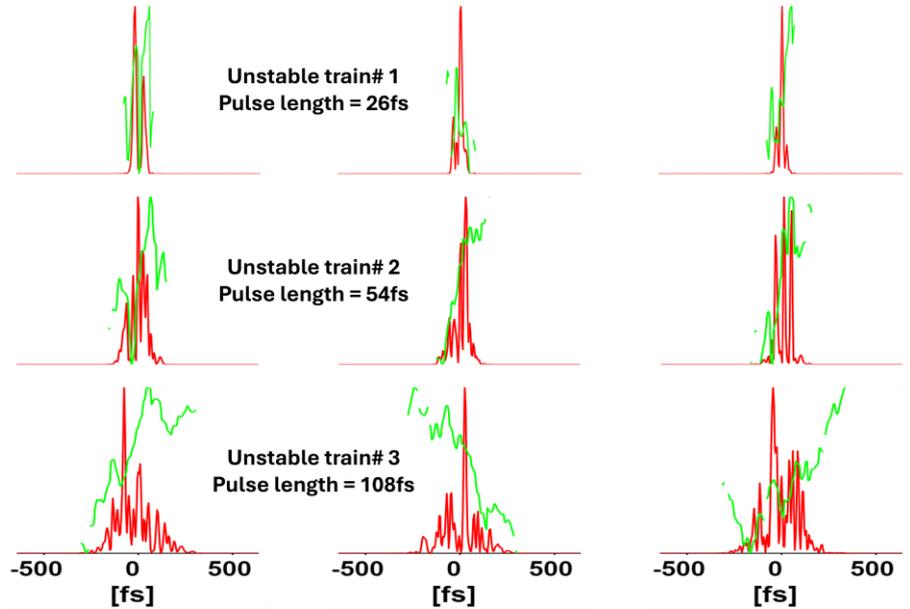

**Fig. 3** Sample pulses from the unstable trains with mean durations of 26, 54, and 108 fs, arranged from top to bottom. The red curves denote the temporal intensities, while the green curves represent the associated temporal phases.

## 4. Results and Discussion

As mentioned earlier, the retrieval process using the RANA approach reliably converges even for an unstable pulse train. Any discrepancy between the retrieved and measured FROG traces therefore reflects random noise and pulse-shape instability, rather than algorithmic stagnation. To quantitatively assess the degree of instability, the *R* value was calculated for both stable and unstable pulse trains. In Fig 4**,** the measured and retrieved SHG FROG traces, along with their corresponding differences, for both stable and unstable pulse trains with a TBP of 10.0, (representing the most complex pulse shape examined in this study) is shown. The difference trace for the stable pulse train exhibits randomly distributed positive and negative pixels with no apparent large-scale connected regions (runs), indicating a high degree of pulse train stability. In contrast, the difference trace for the unstable pulse displays a few large runs, revealing the presence of systematic errors and, consequently, pulse-train instability. Additionally, a distinct vertical feature visible in both the measured and retrieved traces of the unstable pulse train indicate the presence of a coherent artifact, **providing additional qualitative evidence of instability within the pulse ensemble**. Note that this feature does not cause the algorithm to yield an anomalously short pulse because it prevents the trace from corresponding to a single pulse, as is assumed by the algorithm, which hence cannot converge to the measured trace and instead retrieves the best possible representative pulse for that trace. For completeness, Appendix B presents additional examples of the measured, retrieved, and difference traces for both stable and unstable pulse trains with TBP values of 2.5 and 5.0.

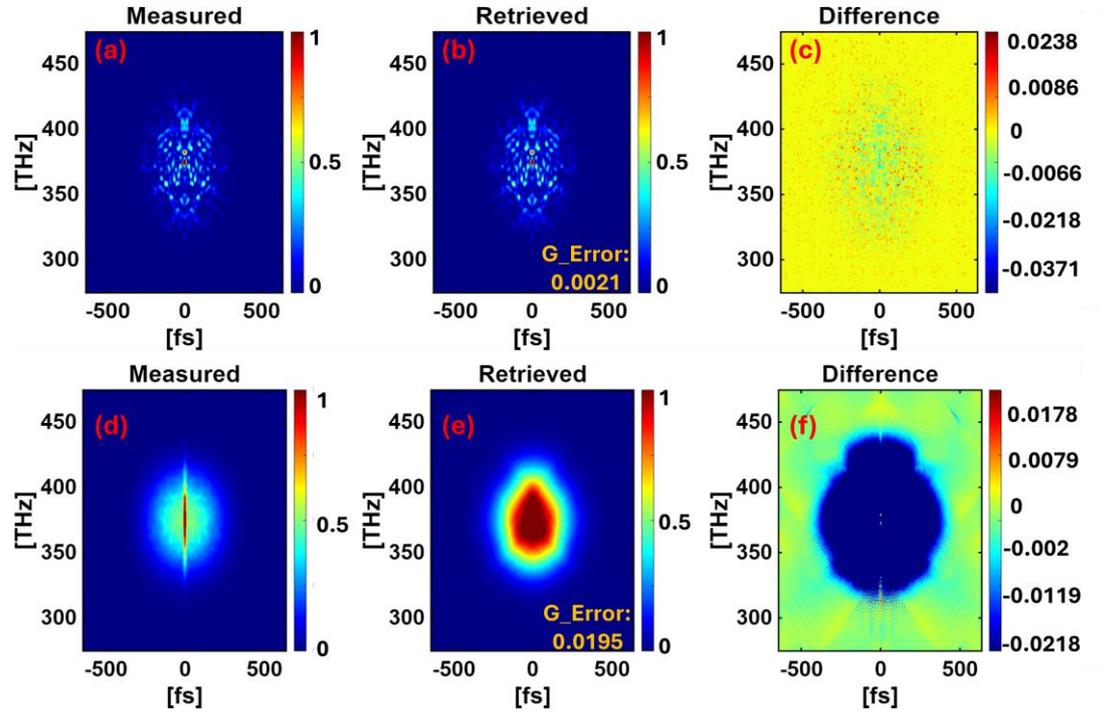

**Fig. 4** Measured and retrieved SHG FROG traces, together with their corresponding difference trace for stable and unstable trains of pulses with a TBP of 10. (a) Measured FROG trace for a stable pulse train. (b) Retrieved FROG trace for the stable pulse train. (c) Difference trace for the stable case, showing randomly signed pixels forming numerous small runs, characteristic of a stable pulse train. (d) Measured FROG trace for the unstable pulse train. (e) Retrieved FROG trace for the unstable pulse train. (f) Difference trace for the unstable case, which exhibits a few large runs, indicating the presence of systematic error and, consequently, pulse-train instability. The FROG trace plots have been normalized to have the same peak values.

We computed the values of $R$ for the measured and retrieved SHG FROG traces for the various pulse trains. Figure 5 shows the weighted-difference traces and 2D runs for both stable and unstable pulse trains with TBP values of 10, 5 and 2.5. As shown, the unstable pulse trains exhibit a few large runs, while the stable pulse trains are characterized by numerous small runs distributed throughout the weighted-difference trace. Our weighted-run approach emphasizes the runs occurring in the central, high-intensity region of the trace, while suppressing the influence of runs in the low-signal trace wings. As we mentioned, this is desirable because the central region contains more physically meaningful information, whereas the edges primarily contain noise or low-intensity artifacts.

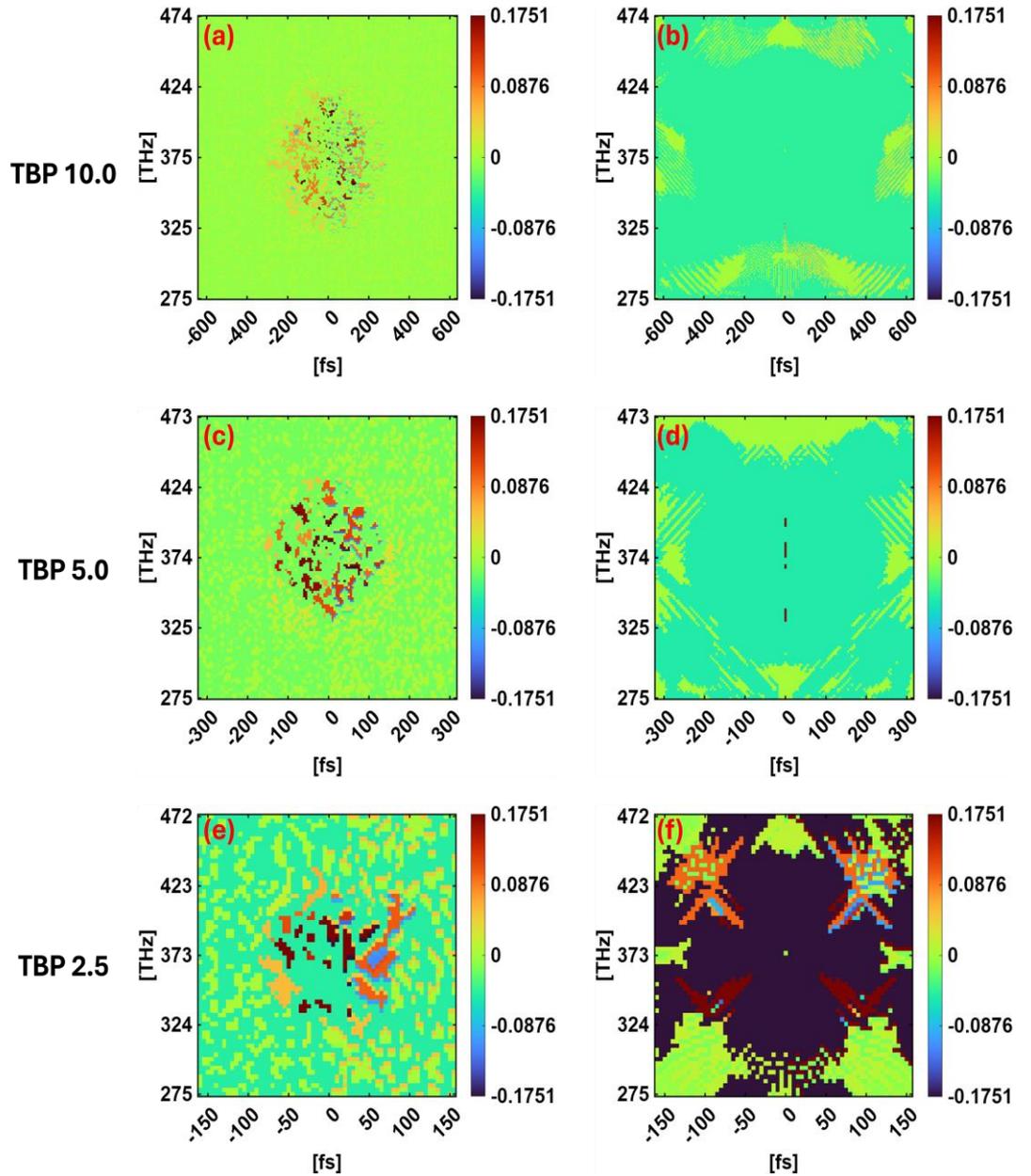

**Fig. 5** The weighted-runs analysis based on the graph method for stable and unstable trains of pulses with TBPs of 2.5, 5, and 10. (a), (c) and (e): the weighted-difference traces of the stable pulse trains each show many small runs, as expected, whereas (b), (d) and (f) show the weighted-difference traces of the unstable pulse trains, each of which exhibits a small number of large runs, an indication of significant systematic error and, consequently, pulse train instability. The larger range of weighted runs in the wings of (f) for the shortest pulse train occurs due to the use of the smallest data array, so that the large values of the measured trace extend over a larger fraction of the data array. In addition, the coherent-artifact spikes in (b) and (d) dominate the weighted-difference traces and so dwarf the remaining runs, which are, as a result, less visible.

For pulse trains with average TBP values of 2.5, 5.0, and 10, respectively, the obtained $R$ values for the stable pulse trains were 0.0361, 0.0362, and 0.0427. In contrast, for the unstable pulse trains, the corresponding $R$ values were much lower, as expected: 0.0051, 0.0052, and 0.0035, respectively. The comparison between the $R$ values obtained for stable and unstable pulse trains is summarized in Table 1. As predicted and discussed in Section 2, stable pulse trains exhibit larger $R$ values than unstable ones. Here, the stable pulse trains exhibit $R$ values that are much larger than those of the corresponding unstable pulse trains. The $R$ parameter thus provides a robust quantitative measure of ultrashort laser pulse-train instability. Note also that $R$ is similar for the different values of TBP.

| Trace Size | TBP | $R_{Stable-train}$ | $R_{Unstable-train}$ |
|---|---|---|---|
| 64 × 64 | 2.5 | 0.0361 | 0.0051 |
| 128 × 128 | 5.0 | 0.0362 | 0.0052 |
| 256 × 256 | 10.0 | 0.0427 | 0.0035 |

**Table 1.** Comparison of the $R$ values for stable and unstable pulse trains. The stable trains consistently exhibit larger $R$ values than the unstable ones, indicating higher pulse-shape stability. This behavior is consistent with the interpretation provided in Section 2. Since the RANA approach is an always-converging retrieval method, any discrepancy between the measured and retrieved traces arises solely from the instability of the pulse trains.

## 5. Conclusions

In this study we introduced a quantitative parameter for assessing ultrashort pulse-train stability by combining reliable multi-shot pulse retrieval using the RANA approach with a 2D-runs analysis applied to the SHG FROG difference trace. Because the RANA algorithm guarantees convergence even in the presence of pulse-train instability and noise, any systematic difference between measured and retrieved traces reflects true pulse-to-pulse variation within the pulse-train, rather than algorithm stagnation. By counting and weighting connected regions of uniform sign (positive or negative) in the difference trace, we define the $R$ value, a metric that distinguishes random, noise-like deviations from systematic discrepancies. Stable pulse trains produce differences dominated by many small, noise-like runs and therefore yield larger $R$ values, whereas unstable trains produce only a few large runs and consequently exhibit much smaller $R$ values.

Using simulated stable and unstable pulse trains with time–bandwidth products ranging from 2.5 to 10, we demonstrated that stable pulse trains consistently yield $R$ values an order of magnitude larger than those of unstable pulse trains. The $R$ value therefore provides a robust and trace-size–independent quantitative measure of pulse-train stability. This approach complements conventional error metrics such as the G-error, which alone cannot distinguish random noise from systematic error and hence physical instability. Our method is readily applicable to a wide range of ultrafast laser systems, from low-power high-repetition-rate oscillators to kHz high-power sources and from the shortest pulses to many-cycle pulses, in all cases where pulse stability is critical.

In future work, we plan to analyze pulse-train instability across a broader range of pulse and stability characteristics. Additionally, this approach could be integrated into a real-time monitoring system for pulse-train stability, in addition to complete-intenisty-and-phase measurement, which would be highly valuable for optimizing ultrashort-pulse lasers in various applications, such as laser-plasma accelerators.

calculations. PA, RJ, and AD wrote and ran runs computation code. PA, AD, BB, and RT wrote the manuscript.

**Funding:** Georgia Tech.

**Data Availability Statement:** Data underlying the results presented in this paper are not publicly available at this time but may be obtained from the authors upon reasonable request.

**Conflicts of Interest:** R.T. owns a company that sells pulse measurement devices. R.J. consults for this company. Other co-authors do not have any conflicts of interest.

## Appendix A – Analytical Proof of the Invariance of the Weighted 2D Runs Statistic $R$

In this appendix, we show that the $R$ parameter is invariant on rescaling a FROG trace, in general, and, in particular, by rescaling it when obeying the discrete Fourier-transform condition, as is often done.

As discussed in the text, a FROG measurement is represented by two real-valued $N \times N$ matrices:

$$I_{FROG}^{meas}(\omega_i, \tau_j) \quad \text{and} \quad I_{FROG}^{retr}(\omega_i, \tau_j), \qquad \text{for } i,j \in \{1, \dots, N\}.$$

The difference trace is $\Delta_{ij} = I_{FROG}^{meas}(\omega_i, \tau_j) - I_{FROG}^{retr}(\omega_i, \tau_j)$, and its sign map

$$S_{ij} = \text{sgn}(\Delta_{ij}) \in \{-1, 0, +1\}.$$

A *2D run* is defined as a maximally 4-connected component of the nonzero entries of $S$ having the same sign. In other words, a run is a contiguous region of pixels where the residual difference has the same sign. Let $\mathcal{R} = \{R_1, \dots, R_K\}$ denote the set of all such runs.

For each run, $R(k)$, which has $R_k$ pixels, we define its mean measured intensity:

$$\bar{I}_k = \frac{1}{|R_k|} \sum_{(i,j) \in R(k)} I_{FROG}^{meas}(\omega_i, \tau_j),$$

And recall that the weighted runs statistic is

$$R = \frac{\sum_{k=1}^{K} \bar{I}_k}{\sum_{i,j} I_{FROG}^{meas}(\omega_i, \tau_j)}.$$

where the denominator represents the total measured intensity of the FROG trace.

*Invariance under Time-Frequency Reparameterization*

We now prove that the statistic $R$ is invariant under any smooth area-preserving reparameterization of the time-frequency domain. Consider the TBP-preserving transformation

$$\tau' = \alpha\tau, \qquad \omega' = \frac{\omega}{\alpha}$$

and define the smooth invertible transformation

$$T(\omega', \tau') = (\alpha\omega', \tau'/\alpha).$$

The Jacobian determinant of this transformation is

$$J = |\det DT| = \left\| \begin{matrix} \alpha & 0 \\ 0 & 1/\alpha \end{matrix} \right\| = 1,$$

Thus, the transformation preserves the phase-space area element

$$d\omega\, d\tau = d\omega'\, d\tau'$$

Under this map,

$$I'^{\text{meas}}_{\text{FROG}}(\omega', \tau') = I^{\text{meas}}_{\text{FROG}}\bigl(T(\omega', \tau')\bigr),$$

and similarly for the retrieved trace. Hence the residual field satisfies

$$\Delta'(\omega', \tau') = \Delta\bigl(T(\omega', \tau')\bigr)$$

Because this transformation $T$, simply rescales the coordinate axes, it does not break apart or merge regions of the residual sign field. As a result, the positive and negative regions of the sign map remain intact, and the number of runs $K$ is unchanged.

Next we will show that $R' = R$, where $R'$ is the $R$ parameter for the transformed coordinates. Let $\Omega_\alpha$ denote a run region in $(\omega, \tau)$ coordinates and $\Omega'_\alpha = T^{-1}(\Omega_\alpha)$ its image. Using the change-of-variables formula and $\det DT = 1$,

$$\int_{\Omega'_\alpha} I'^{\text{meas}}_{\text{FROG}}(\omega', \tau')\, d\omega'\, d\tau' = \int_{\Omega_\alpha} I^{\text{meas}}_{\text{FROG}}(\omega, \tau)\, d\omega\, d\tau,$$

and $|\Omega'_\alpha| = |\Omega_\alpha|$. Hence the mean intensity of each run is preserved,

$$\bar{I}'_k = \bar{I}_k.$$

The total trace intensity is also invariant,

$$\int_{\Omega'} I'^{\text{meas}}_{\text{FROG}} = \int_{\Omega} I^{\text{meas}}_{\text{FROG}}.$$

Therefore, the number of runs, the mean intensity of each run, and the total intensity of the trace are unchanged under the transformation. Consequently,

$$R' = R.$$

The statistic $R$ depends only on the intrinsic geometry of the residual sign field and is invariant under time–bandwidth–preserving reparameterizations of the FROG trace.

## Appendix B – Logarithmic SHG FROG Traces (Measured, Retrieved, and Difference)

This appendix presents the measured, retrieved, and difference SHG-FROG traces for time-bandwidth products (TBP) of 2.5 and 5.0, shown for both stable and unstable pulse trains. Each figure contains three panels: the measured, the retrieved, and the difference traces (measured minus retrieved), displayed on a logarithmic scale to emphasize weak features.

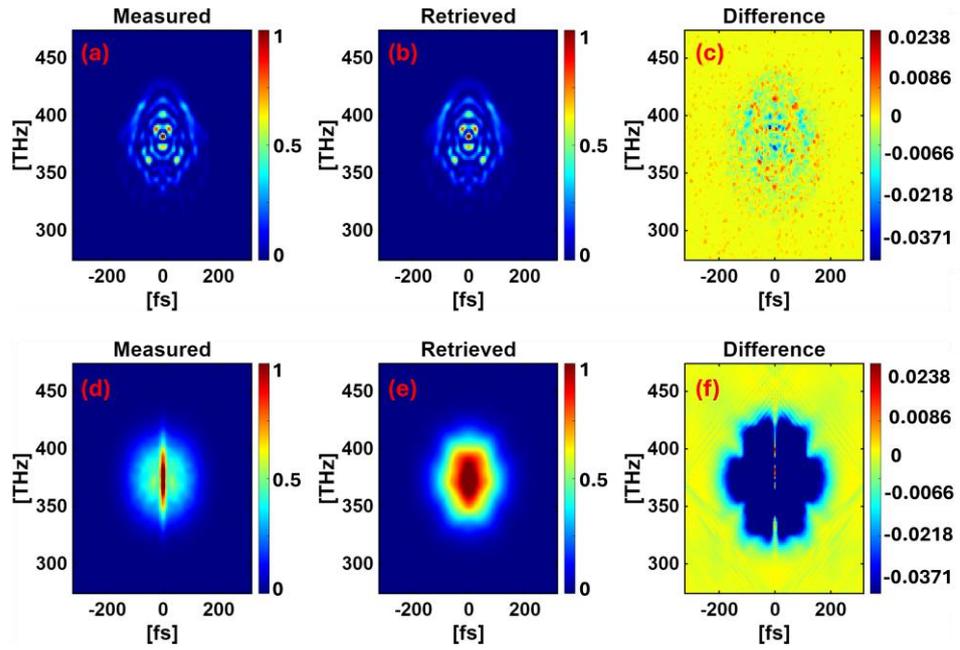

**Figure B1. (a)** Measured, (b) retrieved, and (c) difference SHG FROG traces for a s**table pulse train with a TBP of 5. (d)** Measured, (e) retrieved, and (f) difference SHG FROG traces for an uns**table pulse train with a TBP of 5.**

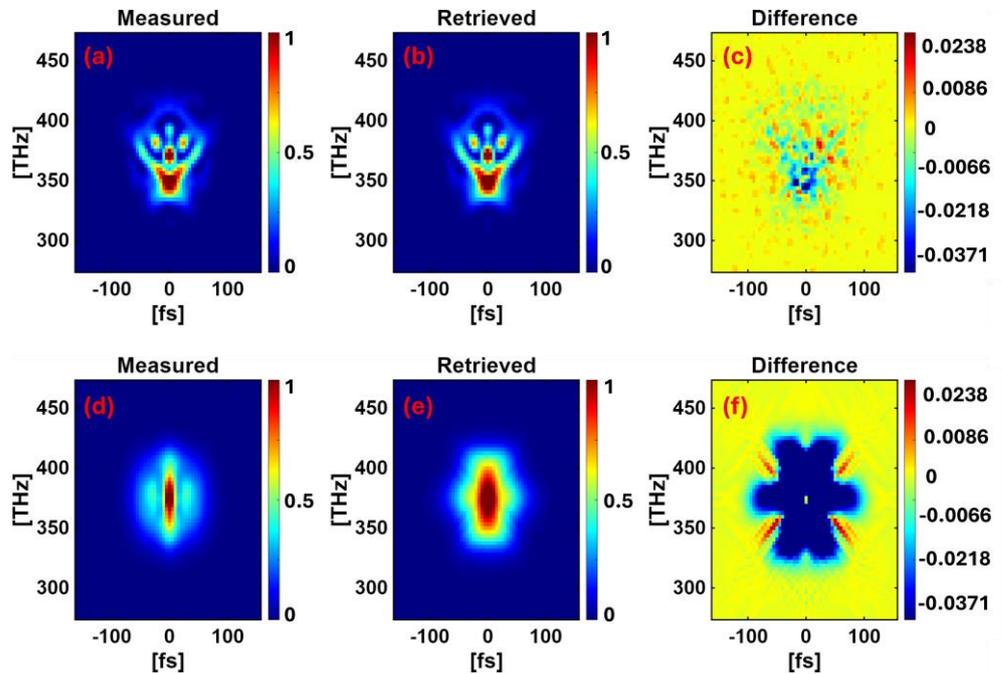

**Figure B2. (a)** Measured, (b) retrieved, and (c) difference SHG FROG traces for a s**table pulse train with a TBP of 2.5. (d)** Measured, (e) retrieved, and (f) difference SHG FROG traces for an uns**table pulse train with a TBP of 2.5.**

Note that each set of figures reveals that same features as the case of TBP = 10 discussed in the main body of the text.